\documentclass[nofootinbib,aps,prl,a4paper,twocolumn,english,superscriptaddress,longbibliography,reprint]{revtex4-2}
\usepackage[mathlines]{lineno}
\usepackage{amsthm}
\usepackage{amsmath}
\usepackage{amssymb}
\usepackage{graphicx}
\usepackage{bm}
\usepackage{color}
\usepackage[dvipsnames]{xcolor}
\usepackage{enumitem}
\usepackage{mathrsfs}
\usepackage{verbatim}
\usepackage{bbold}
\usepackage{braket}
\usepackage{lipsum}
\usepackage[colorlinks, breaklinks, 
linkcolor=OrangeRed,
citecolor=RoyalBlue,
urlcolor=RoyalBlue]{hyperref}

\usepackage{tikz}

\usepackage{graphicx}

\newcommand{\tr}{\text{Tr}}
\newcommand{\ave}[1]{\langle #1 \rangle}
\newcommand{\cl}[1]{\hat{\mathcal{#1}}}

\newcommand{\new}[1]{\textcolor{black}{#1}}
\newcommand{\nnew}[1]{\textcolor{black}{#1}}

\newcommand{\scalemath}[2]{\scalebox{#1}{\begin{math} {#2} \end{math}}}

\begin{document}
\title{Iterative construction of conserved quantities in dissipative nearly integrable systems}
\author{Iris Ul\v{c}akar}
\affiliation{Jo\v{z}ef Stefan Institute, 1000 Ljubljana, Slovenia}
\affiliation{University of Ljubljana, Faculty for physics and mathematics, 1000 Ljubljana, Slovenia}
\author{Zala Lenar\v{c}i\v{c}}
\affiliation{Jo\v{z}ef Stefan Institute, 1000 Ljubljana, Slovenia}

\begin{abstract} 
Integrable systems offer rare examples of solvable many-body problems in the quantum world. Due to the fine-tuned structure, their realization in nature and experiment is never completely accurate, therefore effects of integrability are observed only transiently. \new{One way to overcome this limitation} is to weakly couple nearly integrable systems to baths and driving: these will stabilize integrable effects up to arbitrary time \new{and encode them} in the stationary state approximated by a generalized Gibbs ensemble. However, the description of such driven dissipative nearly integrable models is challenging and no exact analytical methods have been proposed so far. Here, we develop an iterative scheme in which integrability breaking perturbations (baths) determine the conserved quantities that \new{play the leading role in a highly efficient} truncated generalized Gibbs ensemble description. \new{Our scheme paves the way for easier calculations in thermodynamically large systems and can be used to construct unknown conserved quantities.}  
\end{abstract}

\maketitle

\new{Integrable models have played a paramount role in our understanding of nonequilibrium dynamics because, in some cases, their dynamics can be followed exactly.} \nnew{A} modern milestones descriptions of integrable models has been the observation that the steady states reached after a sudden excitation are locally describable with a generalized Gibbs ensemble (GGE) \cite{rigol07,ilievski15a,vidmar16,essler16,caux16}. This observation was a natural generalization of the equilibration of thermalizing generic models, where the steady state is locally described with a thermal Gibbs ensemble \cite{rigol08,polkovnikov11}\nnew{. The} applicability of generalized Gibbs ensembles on a finite timescale was first established experimentally with cold atoms \cite{Langen2015}. \nnew{Recently,} also spatially inhomogeneous dynamics of integrable systems \nnew{was} formulated using a local GGE description and generalized hydrodynamics \cite{castro-alvaredo16,bertini16a,doyon20}, receiving experimental confirmations as well \cite{malvania21,schemmer19,cataldini22,moller21,yang23}. 

Due to their fine-tuned interactions, integrable models cannot be exactly realized in nature. On short to intermediate timescales, integrable models are realizable with quantum simulators \cite{kinoshita06,meinert15,langen15} and can approximately describe some materials \cite{hess07,niesen14,scheie21}. On longer timescales, additional Hamiltonian terms, such as longer range interaction in trapped ions experiment, coupling to phonons in solid state experiment, trapping potential, transverse couplings, and loss of atoms in cold atomic setups prevent integrable effects from \new{persisting up to arbitrary times} \cite{tang17,schemmer19,malvania21,bastianello20}. \nnew{The} only way to sustain integrable effects in nearly integrable systems up to arbitrary times is to drive them out of equilibrium \cite{lange18,lange17,lenarcic18,reiter21,schmitt22}. \nnew{To} prevent heating \new{due to driving, such systems must also be weakly open}. \nnew{Numerical} evidence shows that then the time evolution \cite{lange18} as well as the steady state \cite{lange17,lenarcic18,reiter21,schmitt22} is again approximately described by a GGE. The equations of motion and the steady-state values for the Lagrange parameters associated with the conserved quantities entering the GGE are given deterministically by the integrability breaking perturbations, i.e., by the drive and coupling to the baths. This opens the possibility for GGE engineering \cite{reiter21} and stabilizing potentially technologically useful phenomena. For example, in spin chain materials, approximately described by the Heisenberg model, efficient spin and energy pumping could be realized \cite{lange17}. So far, time-dependent GGE descriptions have also been used to describe the effect of particle loss in cold atoms \cite{bouchoule20,hutsalyuk21} and non-interacting systems coupled to baths \nnew{\cite{reiter21,rossini21,schmitt22,gerbino23,perfetto23,rowlands23}}. 

While the expanded role of GGE \nnew{as a} description of weakly open, nearly integrable systems is fundamentally important, in this case, concrete calculations of dynamics and steady states are much more demanding than in isolated systems. Weak integrability breaking, which causes non-elastic scatterings and a slow reshuffling of quasiparticle content, makes the usual quasiparticle treatment of integrable models much harder; for interacting systems analytically (probably) impossible \cite{bouchoule20,hutsalyuk21}. \new{One possible simplification} is to approximate the GGE with \nnew{macroscopically many Lagrange parameters} with its truncated version. Such an approximation has been used in the context of isolated integrable systems \cite{pozsgay13,fagotti13gge2,essler16,pozsgay17}, as well as in driven dissipative, nearly integrable systems \cite{lange17,lange18,reiter21}. 

\new{We propose a new iterative scheme, which adds the leading conserved quantities to the truncated GGE iteratively,} as suggested by the driving and dissipation itself. \new{Here, leading means having the main contribution to the GGE}. We examine the convergence to the exact result and show that a good approximation is typically achieved within a few steps. \new{\nnew{Meaning}, we have to find a solution of a few coupled equations for a few leading conserved quantities, instead of considering extensively many coupled equations for all conserved quantities.
The only input for the method is the basis from which iterative conserved quantities are constructed.
}

\noindent {\it Setup.}
We consider driven dissipative, nearly integrable setups with the dominant unitary dynamics given by an integrable Hamiltonian $H_0$. Weak driving and dissipation could be due to a Floquet unitary drive and coupling to a (thermal) bath; however, for simplicity, we will consider weak coupling to non-thermal Markovian baths, whose action is described by Lindblad operators $L_i$. As pointed out \nnew{previously} \cite{lange17, lange18,lenarcic18,reiter21,schmitt22}, non-static integrability breaking perturbations should stabilize a GGE generically, and the formalism described below can treat them all. The Liouville equation gives the dynamics of the density matrix operator,
\begin{equation}
\cl{L} \rho = -i [H_0,\rho] + \cl{D} \rho, \ 
 \cl{D} \rho = \epsilon \sum_i L_i \rho L_i^\dagger - \frac{1}{2}\{L_i^\dagger L_i, \rho\},
 \label{eq::liouvillian}
\end{equation}
where $\epsilon\ll 1$ is \nnew{the strength} of the coupling to baths. We will consider homogeneous coupling to baths, where Lindblad operators $L_i$ of the same form act on every site $i$. 

Perturbatively, the zeroth order approximation to the steady state is of a diagonal form in terms of eigenstates of $H_0$, $H_0\ket{m} = E_m \ket{m}$,
\begin{align}\label{eq::diag}
& \lim_{\epsilon\to 0}\lim_{t\to\infty}\rho
= \sum_m a_m \ket{m}\bra{m}
\equiv \rho_D.
\end{align}
Weights $a_m$ are obtained from the kernel of the dissipator projected on the diagonal subspace \cite{lenarcic18, block-diag}, 
$\sum_n D_{mn} a_n = 0$, $D_{mn} = \ave{m| (\cl{D}\ket{n}\bra{n}) | m}$. 
\new{If the dissipator preserves a symmetry of the Hamiltonian, eigenstates can be taken within the symmetry sector with a unique steady state. 
}

As suggested in our previous works \cite{lange17,lange18,reiter21,lenarcic18}, for integrable $H_0$, the zeroth order diagonal ensemble $\rho_D$ should be thermodynamically equivalent to a generalized Gibbs ensemble, 
\begin{equation}\label{eq::GGE}
\lim_{\epsilon\to 0}\lim_{t\to\infty}\lim_{L\to\infty}\rho = \frac{e^{-\sum_m \lambda_m C_m}}{\tr[e^{-\sum_m \lambda_m C_m}]}
\equiv \rho_{\boldsymbol{\lambda}},
\end{equation}
where $C_m$ are the (quasi-)local conserved quantities of the underlying integrable model, $[C_m, H_0]=0$, and $\lambda_m$ the associated Lagrange multipliers. In the steady state, the latter are determined by the stationarity conditions for all conserved quantities
\begin{equation}\label{eq::cond}
\ave{\dot{C}_{m'}} 
= \tr\left[C_{m'} \cl{D} \frac{e^{-\sum_m \lambda_m C_m}}{\tr[e^{-\sum_m \lambda_m C_m}]}\right]
\stackrel{!}{=} 0 \quad \forall m'.
\end{equation}
\new{I.e., one must find the set of $\lambda_m$ for which the flow of all conserved quantities is zero ($\stackrel{!}{=} 0$).}
This equation is very instructive: (i) it tells us that the form of integrability breaking dissipators \nnew{(Lindblad operators)} will determine the $\lambda_m$ values, and (ii) in order to find $\lambda_m$, one must solve a set of coupled non-linear equations. \nnew{To} reduce the complexity of step (ii), an approximate description in terms of a truncated GGE (tGGE) with a finite number of included conserved quantities has been used \cite{lange17,lange18,reiter21}.
\new{In that case, the expectation values of included conserved charges and of local operators constituting them were well captured. However, other local observables showed stronger deviations, particularly those overlapping with quasi-local conserved operators. To partially mend for that, the diagonal part of \nnew{latter observables} was included in the tGGE.}

\nnew{Also} the dynamics towards the steady state can be approximated with a time-dependent GGE \nnew{\cite{lange18}}. The equation of motion for $\lambda_m(t)$ is derived by the use of super-projector $\hat{P}$ onto slow modes, which are for the GGE Ansatz naturally given by the operators $\frac{\partial \rho_{\boldsymbol{\lambda}}}{\partial \lambda_m}$ tangential to the GGE manifold,
\begin{align}\label{eq::P}
\hat{P} X 
&= - \sum_{m,n} 
\frac{\partial \rho_{\boldsymbol{\lambda}}}{\partial \lambda_m} (\chi^{-1})_{m,n} \tr[C_{n}X].
\end{align}
Here, 
$\chi_{m,n} = - \tr[C_m \frac{\partial \rho_{\boldsymbol{\lambda}}}{\partial \lambda_{n}}] = \ave{C_m C_{n}}_{\rho_{\boldsymbol{\lambda}}} - \ave{C_m}_{\rho_{\boldsymbol{\lambda}}} \ave{ C_{n}}_{\rho_{\boldsymbol{\lambda}}}$
is the $\{m,n\}$ entry of matrix $\chi$ and $\ave{O}_{\rho_{\boldsymbol{\lambda}}} = \tr[\rho_{\boldsymbol{\lambda}} O]$.
Applying the super-projector to the slow dynamics on the GGE manifold,
\begin{align}
\hat{P} \dot{\rho}_{\boldsymbol{\lambda}} &=
- \sum_{m,n} 
\frac{\partial \rho_{\boldsymbol{\lambda}}}{\partial \lambda_m} (\chi^{-1})_{m,n} \tr[C_{n}\cl{D} \rho_{\boldsymbol{\lambda}}]
= \sum_{m} 
\frac{\partial \rho_{\boldsymbol{\lambda}}}{\partial \lambda_m} \dot{\lambda}_m
\end{align}
gives the rate of change for the Lagrange multiplier associated with $C_m$,
\begin{equation}\label{eq::flow}
\dot{\lambda}_m = - \sum_{n} (\chi^{-1})_{m,n} \tr[C_{n}\cl{D} \rho_{\boldsymbol{\lambda}}].
\end{equation}
In the super-projector language it is given by the flow along the corresponding tangential direction. Here, the initial conditions $\lambda_m(0)$ are given by the initial state, as in the prethermal state \cite{moeckel08,bertini15}.

{\it Iteratively constructed truncated GGE.}
\nnew{We} use the above super-projector technique to iteratively \new{add the leading conserved quantities to} a truncated description of the steady state for a given dissipator $\cl{D}$. Given that the steady state Lagrange parameters are selected by the dissipator, Eq.~\eqref{eq::cond}, we will, in the first place, use $\cl{D}$ to select the conserved quantities that we include in a truncated GGE Ansatz. If such an iterative truncated description converges to the exact one quickly, the procedure reduces the number of conditions \eqref{eq::cond} that need to be solved.

In the procedure, we iteratively construct conserved quantities $\tilde{C}_k$ from the \new{user-defined operator} basis $Q_m$. \new{The latter} should ideally be the set of \new{all} known (quasi)-local conserved quantities $Q_m=C_m$ of the integrable model $H_0$, \new{but one can also restrict it to contain only some of them. If all conserved quantities are not known or are hard to work with,} one can use the basis with projectors $Q_m = \ket{m} \bra{m}$ onto \new{all} eigenstates of $H_0$ \new{within the symmetry sector with a unique steady state}. \new{However, this introduces certain finite size effects that we discuss in the next section}. 

The iterative procedure has the following steps:

\noindent\underline{\it Step 0:}
Start with a thermal state 
$\rho^{(0)}_{{\boldsymbol{\tilde\lambda}}}  = e^{-\tilde\lambda^{(0)}_0 H_0}/Z$ and find $\tilde\lambda^{(0)}_0$ from condition \eqref{eq::cond}, $\tr[H_0 \cl{D} e^{-\tilde\lambda^{(0)}_0 H_0}]\stackrel{!}{=} 0$.

\noindent\underline{\it Step 1:}
Add the first iterative conserved quantity of the form
\begin{align}
\tilde{C}_1 &= \mathcal{N}_1^{-1}
\sum_m w^{(1)}_m Q_m, \notag\\ 
w^{(1)}_m&= - \sum_n(\chi_{(0)}^{-1})_{mn} \tr[Q_n \cl{D} \rho_{{\boldsymbol{\tilde\lambda}}}^{(0)}], \label{eq::d1}
\end{align} 
where, according to Eq.~\eqref{eq::flow}, weights $w^{(1)}_m$ are given by the flows along additional directions 
$\frac{\partial \rho_{\boldsymbol{\lambda}}}{\partial \lambda_m}|_{\rho^{(0)}_{{\boldsymbol{\tilde\lambda}}}}$ when we allow for a GGE manifold that is not one dimensional (thermal) as in Step 0, but is spanned by additional \new{basis} $Q_m$ conserved quantities. 
A new direction 
$\frac{\partial \rho_{\boldsymbol{\lambda}}}{\partial \lambda_m}|_{\rho^{(0)}_{{\boldsymbol{\tilde\lambda}}}}$ that causes a stronger correction to the existing solution is more important and should be weighted by a stronger bias $w_m^{(1)}$. 
In the end, we find 
$\{\tilde \lambda_0^{(1)}, \tilde \lambda_1^{(1)}\}$ for 
$\rho^{(1)}_{\boldsymbol{\tilde{{\lambda}}}}\propto e^{-\tilde\lambda_0^{(1)} H_0 - \tilde\lambda_1^{(1)} \tilde C_1}$
from the condition \eqref{eq::cond} for $H_0$ and $\tilde{C}_1$.
 
\noindent\underline{\it Step k:}
Add $k$th iterative conserved quantity 
\begin{align}
\tilde{C}_k &= 
\mathcal{N}_k^{-1}
\sum_m w^{(k)}_m Q_m, \notag\\
w^{(k)}_m&= - \sum_n \left(\chi_{(k-1)}^{-1}\right)_{mn} \tr[Q_n \cl{D} \rho_{\boldsymbol{\tilde\lambda}}^{(k-1)}] \label{eq::weight}
\end{align}
and find 
$\{\tilde\lambda_{k'}^{(k)}\}$
for 
$\rho_{\boldsymbol{\tilde{\lambda}}}^{(k)}\propto e^{-\sum_{k'=0}^k \tilde\lambda_{k'}^{(k)} \tilde{C}_{k'}}$
from the set of $k+1$ conditions \eqref{eq::cond} for $\{\tilde{C}_{k'}\}_{k'=0}^k$, where we denote the $\tilde{C}_0 = H_0$.
\nnew{Normalization $\mathcal{N}_k$ can be absorbed into the corresponding Lagrange parameter}. However, it can also be chosen such that $\tilde{C}_k$ scales \new{as an extensive operator}.
The susceptibility matrix,
$(\chi_{(k)})_{m,n} = \ave{Q_m Q_{n}}_{\rho_{\boldsymbol{\tilde\lambda}}^{(k)}} - \ave{Q_m}_{\rho_{\boldsymbol{\tilde\lambda}}^{(k)}} \ave{ Q_{n}}_{\rho_{\boldsymbol{\tilde\lambda}}^{(k)}}$,
must be evaluated in each iterative step.
Matrix $\chi_{(k)}$ is not invertible for the non-local basis with $Q_m=\ket{m}\bra{m}$, however, one can regularize it as explained in the Supplementary material (SM)\nnew{, \cite{supmat}}.
In case of (additional) unitary (Floquet) perturbations, the iterative procedure can be generalized to cover those as well \nnew{\cite{supmat}}.

{\it Results.}
\new{First, to quantify the (finite-size) error of different (truncated) GGE descriptions on length-scale $\ell$}, we use the distance between density matrices \cite{fagotti13gge2} 
\begin{equation}\label{eq::dist}
d\left(\rho_{1}, \rho_{2}\right)=\sqrt{\frac{\operatorname{Tr}[\left(\rho_{1}-\rho_{2}\right)^{2}]}{\operatorname{Tr}[\rho_{1}^{2}]+\operatorname{Tr}[\rho_{2}^{2}]}}
\end{equation}
and compare reduced density matrices on $\ell$ consecutive sites for different (truncated) GGE descriptions $\rho_{\boldsymbol{\lambda},\ell}$ and the diagonal solution $\rho_{D, \ell}$, Eq.~\eqref{eq::diag}.
As noted in Ref. \cite{fagotti13gge2}, the distance between two reduced space GGEs scales as 
$d(\rho_{\boldsymbol{\lambda},\ell}, \rho_{\boldsymbol{\lambda}',\ell})
\sim \ell \sum_i|\lambda_i - \lambda'_i|
$ for large enough $\ell$. Also for $\rho_D$ close to a GGE, we expect $d(\rho_{\boldsymbol{\lambda},\ell},\rho_{D,\ell})$ to scale with $\ell$ for large enough $\ell$.
\begin{figure}
\centerline{\includegraphics[width=\columnwidth]{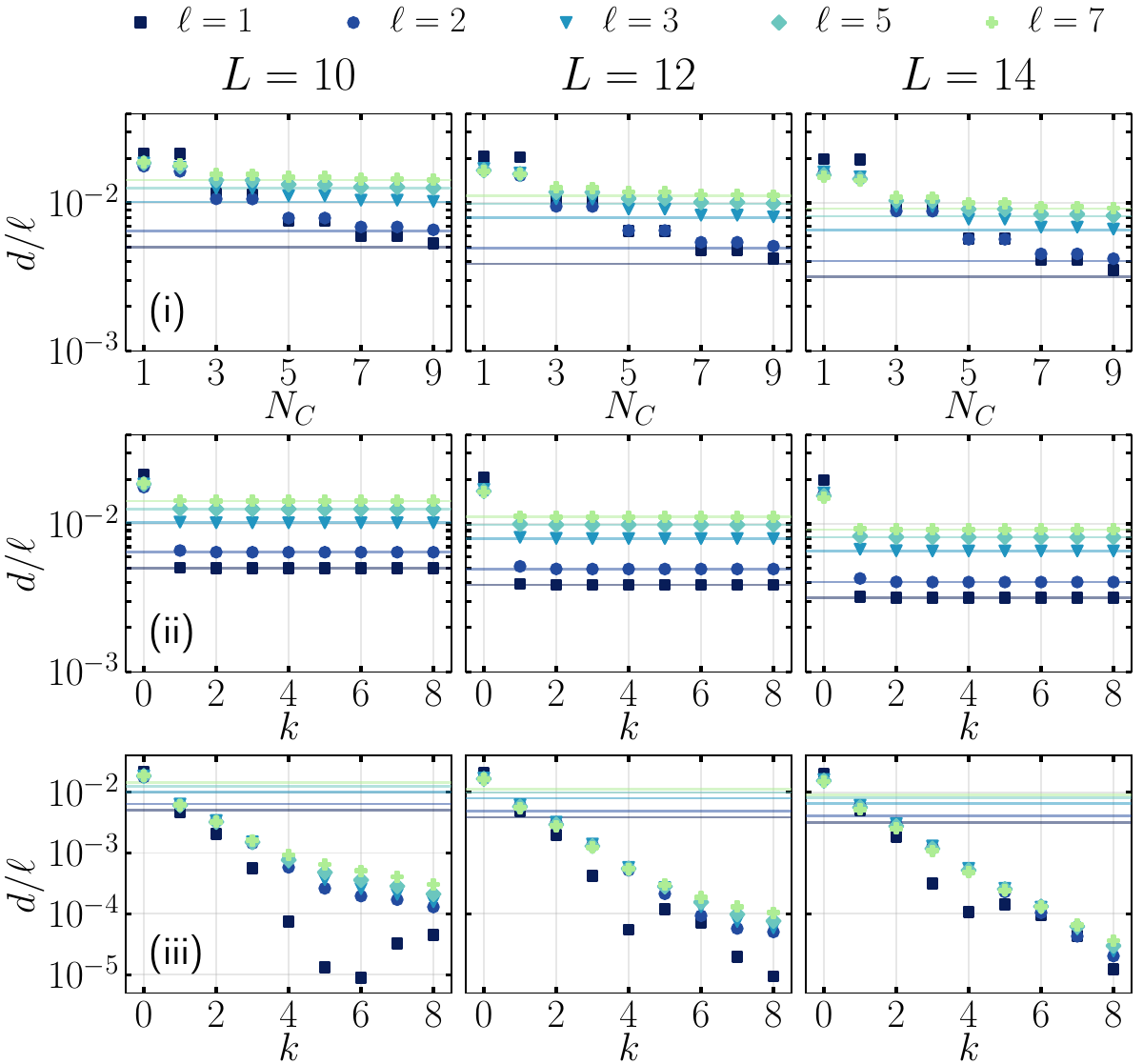}}
\caption{Scaled distances $d(\rho_{\boldsymbol{\lambda},\ell},\rho_{D,\ell})/\ell$, Eq.~\eqref{eq::dist}, between the reduced density matrices of support $\ell$ for the diagonal steady state ensemble, Eq.~\eqref{eq::diag}, and different (truncated) generalized Gibbs ensembles. Solid lines correspond to the best GGEs with all $N_C^{\text{all}}=2L-2$ local conserved quantities on a given system size. From top to bottom, we consider tGGEs with: (i)~$N_C$ local conserved quantities $C_m, m=0,...,N_C-1$, 
(ii)~iterative scheme after the $k$th step for the basis $Q_m=C_m$, $m=1,...,N_C^{\text{all}}-1$, (iii)~iterative scheme after the $k$th step for basis $Q_m=\ket{m}\bra{m}$. Parameters: Ising model with $h_x=0.6$, $J=1$, system sizes $L=10,12,14$.}
\label{fig1}
\end{figure}

In Fig.~\ref{fig1}, we show the scaled distances $d(\rho_{\boldsymbol{\lambda},\ell},\rho_{D,\ell})/\ell$ for our first example, the transverse field Ising model 
\begin{equation}
H_0 = \sum_i J \sigma^z_i \sigma ^z_{i+1} + h_x \sigma ^x_i,    
\end{equation}
which is a paradigmatic non-interacting integrable model. It preserves a series of local extensive operators $C_m$ \cite{grady82}, \nnew{see \cite{supmat}}. We choose the following Lindblad operators \new{of a rather general form,} with $a=0.2$,
\begin{equation} 
L_i = S_i^+S_{i+1}^- + i S_{i+1}^-S_{i+2}^+ + a \sigma_i^x \sigma_{i+1}^z, \label{eq::isingL}
\end{equation}
\new{which stabilize a non-trivial steady state.}
\nnew{Such structured dissipators could be realized with trapped ions simulators \cite{barreiro11,reiter21} or superconducting circuits \cite{mi23}. Moreover, \nnew{we stress} that our algorithm is generic and applicable \nnew{to any non-hermitian} Lindblad operators leading to (within the symmetry sector) a unique, non-trivial steady state.}

\begin{figure}[t!]
\includegraphics[width=\columnwidth]{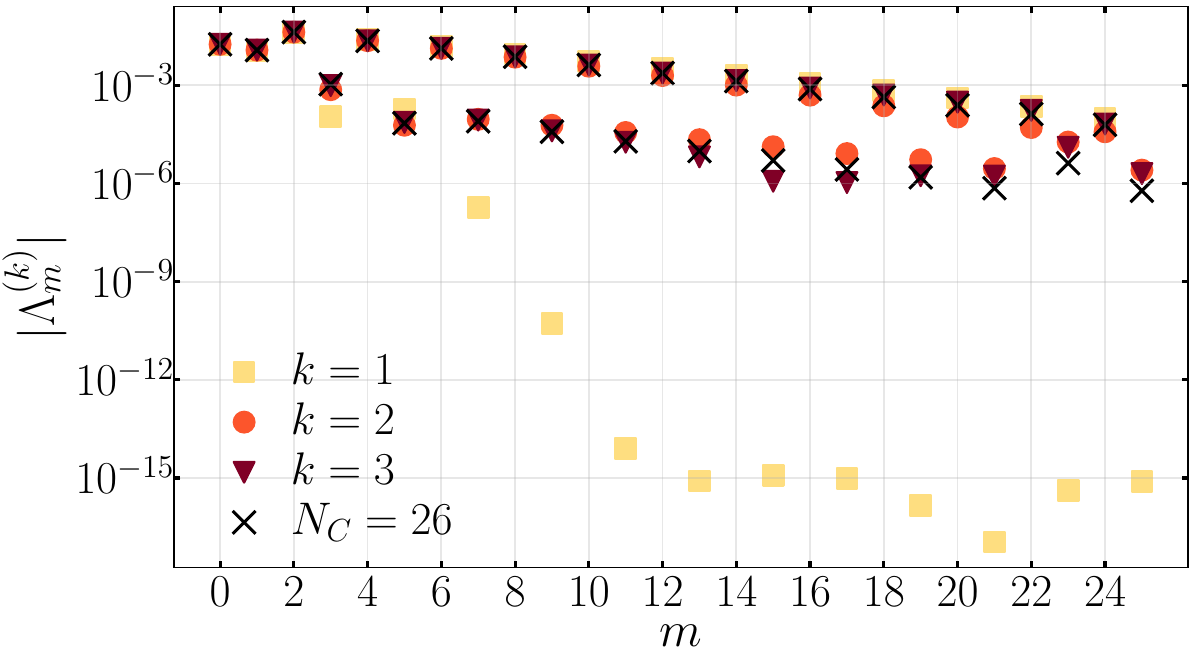}
\caption{
\new{Interpreting the iterative procedure using the basis with all local conserved quantities $Q_m=C_m, m=1,...,2L-3$.
The weights 
$\Lambda_m^{(k)}=|\sum_{k'=1}^k \tilde{\lambda}_{k'}^{(k)} w_m^{(k')}/\mathcal{N}_{k'}|$ at each basis element $Q_m$ after $k=1,2,3$ iterative steps show which basis elements are more important. The exact weights are shown with crosses. Parameters: transverse field Ising model with $h_x=0.6$ and $J=1$ on $L=14$ sites.
}}
\label{fig2}
\end{figure}

\new{Solid lines on all  Fig.~\ref{fig1} panels denote the saturated distances $d_{s,\ell}$ for the best GGE solution on a given system size $L$, including all $N_C^{\text{all}}=2L-2$ local conserved quantities on that system size. The solution is obtained \nnew{by solving} extensively many $N_C^{\text{all}}\sim 2L$ coupled Eq.~\eqref{eq::cond}.
As $L$ is increased, saturated $d_{s,\ell}$ are decreased because in the thermodynamic limit, the complete GGE and the diagonal solution are equivalent, \new{$\lim_{L\to\infty}d_{s,\ell}\to 0$}. Finite $d_{s,\ell}$ are just a finite-size effect.}
(i)~\nnew{Traditional tGGE: }In the first row, we show the convergence to $d_{s,\ell}$ for \nnew{a tGGE} based on $N_C \le N_C^{\text{all}}$ most local conserved quantities $C_m, m = 0, \dots N_C-1$, where $C_0$ is the Hamiltonian. To find the solution, $N_C$ coupled Eqs.~\eqref{eq::cond} are solved. As $N_C$ is increased, a better description with smaller distances $d$ is obtained, 
\new{but convergence is rather slow.} 
(ii)~\nnew{Iterative tGGE with local basis: }In the middle row, we show the convergence of $d(\rho^{(k)}_{\boldsymbol{\tilde\lambda},\ell},\rho_{D,\ell})/\ell$ with respect to the number of iterative steps $k$ for the iterative scheme using the basis of \nnew{{\it all} local conserved quantities $Q_m=C_m$, excluding the Hamiltonian.}
The saturated distances $d_{s,\ell}$ (solid lines) are approached after only $k=1$ iterative step. The advantage is that we now solve for $k+1=2$ instead of \new{extensively many} $N_C^{\text{all}} \sim 2L$ conditions \eqref{eq::cond} in order to find the steady state.
(iii)~\nnew{Iterative tGGE with non-local basis: }In the last row, we present the convergence of the iterative scheme in the non-local basis $Q_m=\ket{m}\bra{m}$ \nnew{of projectors onto the eigenstates of $H_0$}. The saturated distances $d_{s,\ell}$ (solid lines) are met after only $k=1$ iterative step. Further reduction of $d$ is largely due to non-local contributions in the non-local basis \nnew{$Q_m$}, \new{which inherently share finite size effects of the diagonal solution $\rho_D$}, Eq.~\eqref{eq::diag}.
\new{In that sense, $\rho_D$ and the iterative solution in the diagonal basis, contain thermodynamically irrelevant information. We should note also that the iterative procedure in the non-local basis cannot be more efficient than finding $\rho_D$. However, its usefulness is in the interpretability, which is achieved by analyzing the structure of the leading conserved quantities. Below, we discuss the advantages of our iterative procedure in both bases.}

\new{When all (quasi-)local conserved quantities are known, the advantage of the iterative procedure performed in their basis is in efficiently interpreting the steady state by establishing the weights $\Lambda_m^{(k)}$ at different basis elements $Q_m=C_m$, \nnew{in Fig.~\ref{fig2} shown for $k=1,2,3$ iterative steps \cite{normlambda}}.
Weights $\Lambda_m^{(k)}$ reveal that conserved quantities $C_{2n}$, which are even under the parity transformation, \nnew{are} more important for our example and are well estimated after a single $k=1$ iterative step. In the next iterative steps, smaller weights at less important odd conserved quantities $C_{2n -1}$ are also captured. To further illustrate the fast convergence, we compare the iterative results to the asymptotic weights (absolute values of Lagrange parameters $|\lambda_m|$ denoted by crosses), obtained by solving \nnew{extensively many conditions \eqref{eq::cond} for all local conserved quantities}. 
}

\begin{figure}[t!]
\includegraphics[width=\columnwidth]{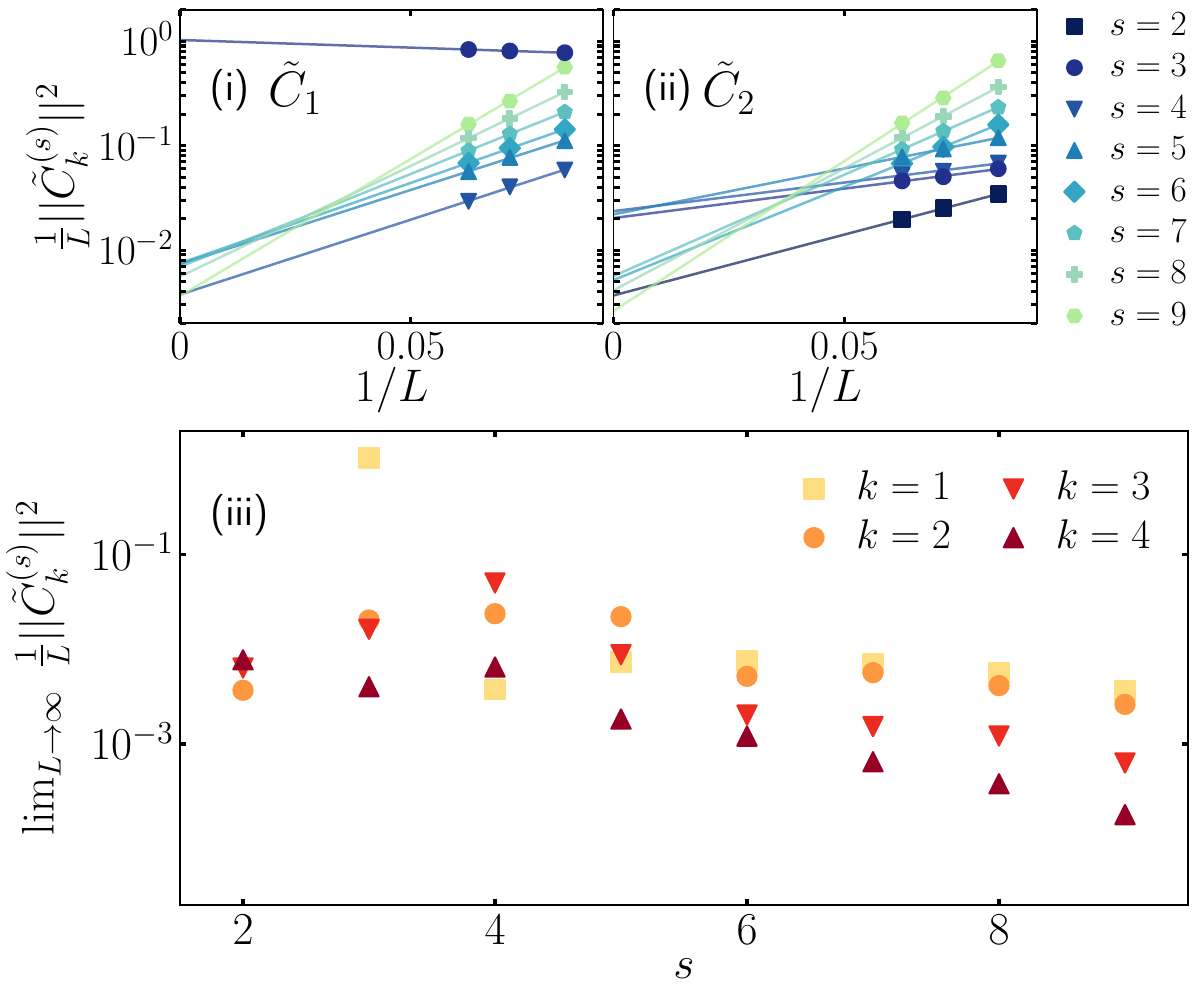}
\caption{
\new{
Nature of iterative conserved quantities, constructed in the basis of projectors on eigenstates, $Q_m=\ket{m}\bra{m}$.
(i,ii)~Finite size scaling of norms $\frac{1}{L}||\tilde{C}_k^{(s)}||^2$ for the part $\tilde{C}_k^{(s)}$ of $k=1,2$ iterative conserved operator $\tilde{C}_k = \sum_s \tilde{C}_k^{(s)}$ that acts non-trivially only on support $s$.
(iii)~Thermodynamic norms $\lim_{L\to \infty}\frac{1}{L}||\tilde{C}_k^{(s)}||^2$ indicate that iterative conserved operators $\tilde{C}_k$ have a quasi-local nature. Parameters: Heisenberg model on $L=12,14,16$.
}}
\label{fig3}
\end{figure}

\new{If conserved quantities are unknown or hard to work with, one should perform the iterative procedure in the non-local diagonal basis $Q_m=\ket{m}\bra{m}$. The usefulness is again in the interpretability, obtained by analyzing the structure of leading \nnew{iterative} conserved quantities. Here, we perform it for the Heisenberg model, 
\begin{equation}
H_0 = \sum_i \sigma^x_i \sigma^x_{i+1} + \sigma^y_i \sigma^y_{i+1} + \sigma^z_i \sigma^z_{i+1},
\end{equation}
known to have additional exotic quasi-local conserved quantities \cite{ilievski15,ilievski15a,ilievski16}. We again use Lindblad operators Eq.~\eqref{eq::isingL}, now with $a=0$ so that we can work in a single (largest) $S^z=0$ magnetization sector. 
An analysis similar to Fig.~\ref{fig1} is performed in the SM, here we focus on the interpretability.
In order to assess the nature \nnew{(i.e. locality)} of iterative conserved quantities, we extract the norm $\lim_{L\to \infty}\frac{1}{L}||\tilde{C}_k^{(s)}||^2$ of the \nnew{part of} iterative conserved quantities \nnew{that acts non-trivially \nnew{on $s$ consecutive sites}, $\tilde{C}_k = \sum_s \tilde{C}_k^{(s)}$}. Operator $\tilde{C}_k^{(s)}$ is obtained by summing the overlaps of $\tilde{C}_k$ with all Pauli strings acting non-trivially between $s$ sites.
In Fig.~\ref{fig3}(i,ii), we plot the norms $\frac{1}{L}||\tilde{C}_k^{(s)}||^2$ for $k=1,2$ and different supports $s$ on given system sizes $L=12,14,16$ \nnew{\cite{heis}.}
\nnew{For small} $L$, norms even increase with $s$. Only when we extrapolate the finite size result to the thermodynamic limit via $1/L$ scaling, Fig.~\ref{fig3}(i,ii), for large enough $s$, norms are decaying exponentially with $s$, Fig.~\ref{fig3}(iii), as expected for quasi-local conserved operators \cite{ilievski15,ilievski16}. 
Thus, only once we remove the finite-size non-local contributions, inherent to the non-local basis, we can conclude that our iterative conserved quantities are a quasi-local superposition of local \cite{grabowski95} and quasi-local conserved quantities of the Heisenberg model \cite{ilievski15,ilievski16}. }

{\it Conclusions and Outlook.}
In this Letter, we propose an iterative construction of \new{conserved quantities of leading importance} to describe nearly integrable, driven dissipative systems. Our approach is motivated by the fact that in \new{such setups}, the integrability breaking perturbations (couplings to baths and drives) determine the \new{Lagrange parameters of a GGE approximation to the steady state}. Here, we use the dissipator to select the combination of conserved quantities contributing significantly to the truncated GGE description. Such a physically motivated construction of truncated GGEs reduces the complexity of calculating steady state parameters, i.e., the Lagrange parameters. Namely, instead of solving \new{extensively} many coupled conditions \eqref{eq::cond} for all (quasi)local conserved quantities, we need to solve for just a few. 
\new{A precise number is model and precision-dependent but is generally expected to be $\mathcal{O}(1)$ and low.}

\new{A clear usefulness of our approach that we already showcased here is in the interpretability: (i) If working in the basis of all local conserved operators, their weights reveal which of them are important for given Lindblads.  (ii) If all (quasi-)local conserved quantities are not known or hard to work with, the iterative procedure can be performed in the non-local basis of projectors on eigenstates. The (quasi-)local structure of leading iterative conserved quantities can be analyzed a posteriori, giving the information about potentially unknown conserved quantities of the unperturbed model \cite{bentsen19}. }

When evaluating the actual reduction of complexity, one should also consider the complexity of evaluating the rate equations \eqref{eq::cond} and building the conserved quantities, Eq.~\eqref{eq::weight}. 
\nnew{Our current study used exact diagonalization, at a gain of thorough benchmarking against the diagonal ensemble, but at the loss of diagonalization itself representing the bottleneck of the procedure.}
At least for non-interacting many-body integrable systems, one can evaluate Eq.~\eqref{eq::cond} and construct iterative conserved quantities~\eqref{eq::weight} in the basis of mode occupation operators with \nnew{polynomial complexity in system size, in which case our method reduces the exponent and simplifies thermodynamically large calculations; see comment \cite{complex} for scaling arguments.} For interacting integrable models, Eqs.(\ref{eq::cond},\ref{eq::weight}) could be evaluated using partition function approach \cite{ilievski16b,ilievski2019}. However, we leave exploring thermodynamic aspects for non-interacting and interacting models to a future study.

\begin{acknowledgments}
We thank A. Rosch and E. Ilievski for several useful discussions. We acknowledge the support by the projects J1-2463, N1-0318 and P1-0044 program of the Slovenian Research Agency, the QuantERA grant QuSiED by MVZI, QuantERA II JTC 2021, and ERC StG 2022 project DrumS, Grant Agreement 101077265. ED calculations were performed at the cluster `spinon' of JSI, Ljubljana. 
\end{acknowledgments}

\bibliography{iterative_iris.bib}

\begin{thebibliography}{55}
\bibitem [{blo()}]{block-diag}%
  \BibitemOpen
  \href@noop {} {\bibinfo {title} {In the case of degeneracies of ${H}_0$, {E}q.~\eqref{eq::diag} should be extended to a block-diagonal form; however, we expect that thermodynamically, both descriptions are equivalent, and we will work with diagonal ensemble here.}}\BibitemShut {Stop}%
\bibitem [{nor()}]{normlambda}%
  \BibitemOpen
  \href@noop {} {\bibinfo {title} {{Here, we used the natural normalization $||\tilde{C}_k||^2 = L$ with $||A||^2 = \frac{\tr[A A^\dagger]}{N_H} - \frac{\tr[A] \tr[A^\dagger]}{N_H^2}$ and $N_H$ the Hamiltonian Hilbert space dimension. The Hamiltonian has it's own Lagrange multiplier, so $\Lambda_0^{(k)}= |\tilde{\lambda}_0^{(k)}|$.}}}\BibitemShut {Stop}%
\bibitem [{hei()}]{heis}%
  \BibitemOpen
  \href@noop {} {\bibinfo {title} {{Here, we use the normalization $\mathcal{N}_k = \sqrt{\frac{1}{L N_H} \sum_m |w_m^{(k)}|^2 }$, where $N_H$ is the Hamiltonian Hilbert space dimension. Therefore $\tilde{C}_k$ are scaling extensively, $||\tilde{C}_k||^2\sim L$, where $||A||^2 = \frac{\tr[A A^\dagger]}{N_H} - \frac{\tr[A] \tr[A^\dagger]}{N_H^2}$}}}\BibitemShut {NoStop}%
\bibitem [{com()}]{complex}%
  \BibitemOpen
  \href@noop {} {\bibinfo {title} {{Our current study used exact diagonalization on Hilbert space of dimension $N_H$ with its own complexity $\mathcal{O}(N_H^3)$. Then the reduction of complexity from $\mathcal{O}(L^2\, N_H)$ to $\mathcal{O}(k^3\, N_H)$ for solving equations~\eqref{eq::cond} does not remove the bottleneck. Here, we used that $k$ is the number of necessary iterative steps and that finding roots for $n$ variables scales as $\mathcal{O}(n^a)$, with $a=2$ for the Powell method employed \cite{press07}. Scaling $\mathcal{O}(k^3N_H)$ is obtained by taking into account that in each (intermediate) iterative step $k'\leq k$, $\mathcal{O}(k'^2)$ root findings are performed, summing up to $\mathcal{O}(k^3)$ in total. However, at least for non-interacting many-body integrable systems, one can evaluate Eq.~\eqref{eq::cond} and construct iterative conserved quantities~\eqref{eq::weight} in the basis of mode occupation operators in $\mathcal{O}(L^b)$ steps. Then our iterative method diminishes the complexity
  from $O(L^{b+2})$ for finding Lagrange multipliers of all conserved quantities to $O(k^3 \, L^b)$ for k iterative conserved} quantities.}}\BibitemShut {Stop}%
\end{thebibliography}

\newpage
\phantom{a}
\newpage

\renewcommand{\thetable}{S\arabic{table}}
\renewcommand{\thefigure}{S\arabic{figure}}
\renewcommand{\theequation}{S\arabic{equation}}
\renewcommand{\thepage}{S\arabic{page}}

\renewcommand{\thesection}{S\arabic{section}}

\onecolumngrid

\title{Iterative construction of conserved quantities in dissipative nearly integrable systems}
\author{Iris Ul\v{c}akar}
\affiliation{Jo\v{z}ef Stefan Institute, 1000 Ljubljana, Slovenia}
\affiliation{University of Ljubljana, Faculty for physics and mathematics, 1000 Ljubljana, Slovenia}
\author{Zala Lenar\v{c}i\v{c}}
\affiliation{Jo\v{z}ef Stefan Institute, 1000 Ljubljana, Slovenia}

\setcounter{figure}{1}
\setcounter{equation}{1}
\setcounter{page}{1}

\begin{center}
{\large \bf Supplemental Material:\\
Iterative construction of conserved quantities in dissipative nearly integrable systems}\\
\vspace{0.3cm}
Iris Ul\v cakar$^{1,2}$ and Zala Lenar\v ci\v c$^{2}$\\
$^1${\it Faculty for Mathematics and Physics, University of Ljubljana, Jadranska ulica 19, 1000 Ljubljana, Slovenia} \\
$^2${\it Department of Theoretical Physics, J. Stefan Institute, SI-1000 Ljubljana, Slovenia} \\
\end{center}

In the Supplemental Material, we (i) show how to regularize the inverse of susceptibility matrix $\chi^{-1}$ in the case of non-local basis of projectors $Q_m=\ket{m}\bra{m}$, (ii) write the explicit form for the local conserved quantities of the transverse field Ising model, \new{(iii) perform the analysis of different tGGE descriptions in the case of the Heisenberg model, and (iv) show how additional unitary perturbation can be included in the iterative procedure.}

\vspace{0.6cm}

\twocolumngrid

\label{pagesupp}

\subsection{Regularization of $\chi^{-1}$ in the non-local basis} \label{app4}

Here we address the invertibility of the susceptibility matrix $\chi$, appearing in the calculation of weights $w_m^{(k)}$ at basis elements $Q_m$ in the iterative conserved quantities $\tilde{C}_k$, Eq.~\eqref{eq::weight},
\begin{align}\label{eq::SMweight}
\tilde{C}_k &= 
\mathcal{N}_k^{-1}
\sum_m w^{(k)}_m Q_m, \notag\\
w^{(k)}_m&= - \sum_n (\chi_{(k-1)}^{-1})_{mn} \tr[Q_n \cl{D} \rho_{\boldsymbol{\tilde\lambda}}^{(k-1)}]
\end{align}
for different choices of operator basis $Q_m$. 
Let us first rewrite the equation for weights $w_m^{(k)}$, Eq.~\eqref{eq::SMweight}, in a vector form. To avoid the cluttering of notation, we omit the step index $k$ in the vector representation. We denote the vector of weights with $\ket{w} = \ket{w_1^{(k)}, \dots w_{M}^{(k)}}$, with $n$th element being the weight at $Q_n$. The vector index runs from $1$ up to the size of the $Q_m$ basis $M$. Additionally we denote with $\ket{q} = \ket{q_1^{(k-1)}, \dots q_{M}^{(k-1)}}$ the vector with entries $q_n^{(k-1)}  = \tr[Q_n \cl{D} \rho_{\boldsymbol{\tilde\lambda}}^{(k-1)}]$. Matrix $\chi$ has size $M \times M$ and it's elements are the connected correlation functions between the basis elements $Q_m$, 
$\chi_{mn} = \tr[Q_m Q_{n} \rho_{{\boldsymbol{\tilde\lambda}}}^{(k-1)}] - \tr[Q_m \rho_{{\boldsymbol{\tilde\lambda}}}^{(k-1)}] \, \tr[Q_{n} \rho_{{\boldsymbol{\tilde\lambda}}}^{(k-1)}]$.
The equation \eqref{eq::SMweight} for weights is then written as
\begin{equation}
    \ket{w} = \chi^{-1}\ket{q}.
    \label{eq::S3}
\end{equation}
and can be solved if $\chi$ is invertible, which is the case for the basis of local conserved quantities $Q_m = C_m$. In the remaining part of this section, we show that for the non-local basis of projectors onto eigenstates of $H_0$, $Q_m=\ket{m}\bra{m}$, matrix $\chi$ is non-invertible and we demonstrate how to calculate weights $\ket{w}$ by regularizing $\chi^{-1}$.

For the non-local $Q_m=\ket{m}\bra{m}$, matrix elements of $\chi$ can be expressed with coefficients of $\rho_{\boldsymbol{\tilde\lambda}}^{(k-1)}$ in the diagonal ensemble $a_m = \bra{m}\rho_{\boldsymbol{\tilde\lambda}}^{(k-1)}\ket{m}$, such that $\chi_{mn} = a_m (\delta_{mn} - a_n)$. Due to the normalization of the density matrix $\tr[\rho_{\boldsymbol{\tilde\lambda}}^{(k-1)}] = 1$, $N_H-1$ coefficients $a_m$ are free parameters, while one depends on the others. 
Here, $N_H$ is the Hilbert space dimension and also the number of basis elements $M = N_H$.
This implies that matrix $\chi$ has $N_H-1$ linearly independent rows (or columns), and therefore rank $N_H-1$. One vector is in the kernel of $\chi$, meaning $\chi$ is non-invertible. The kernel vector $\ket{v}$ is in the eigenbasis of $H_0$ of form $\ket{v} = \frac{1}{\sqrt{N_H}}\ket{1, \dots, 1}$, since the sum of any row (as well as column) in matrix $\chi$ in this basis equals zero, 
\begin{equation}
    \chi\ket{1, \dots, 1} = 0, \qquad \bra{1, \dots, 1} \chi = 0.
    \label{eq::S4}
\end{equation}
Note also that $\tr[\cl{D} \rho_{\boldsymbol{\tilde\lambda}}^{(k-1)}] = \sum_n q_n = 0$, or 
\begin{equation}
    \braket{1, \dots, 1|q} = 0
    \label{eq::S5}
\end{equation}
due to the dissipator $\cl{D}$ in Eq.~\eqref{eq::liouvillian} being of Lindblad form.

To avoid the issue of invertibility of $\chi$, one can instead of Eq.~\eqref{eq::S3} solve 
\begin{equation}
    \chi \ket{w} = \ket{q}, \quad \ket{w}=\ket{w'} + c\ket{v}
    \label{eq::S6}
\end{equation}
with a set of solutions parametrized by $c \in \mathbb{C}$,  since $\chi (\ket{w'} + c\ket{v}) = \chi \ket{w'} = \ket{q}$. Below we show that
\begin{equation}
    \ket{w'} = \tilde\chi^{-1} \ket{q}, \qquad \tilde\chi = \chi + \ket{v}\bra{v}
    \label{eq::regInv}
\end{equation}
is a solution giving rise to a traceless iterative conserved quantity $\tr[\tilde{C}_k] = 0$ and that the choice of factor $c$ does not affect the iterative steady state density matrix $\rho^{(k)}_{\boldsymbol{\tilde\lambda}}$.

In Eq.~\eqref{eq::regInv} we introduced a regularized inverse $\tilde\chi^{-1}$, motivated by the fact that we can choose an orthonormal basis, where one basis vector is the kernel vector $\ket{v}$ (subspace $\tilde{Q}$) while others form a complementary subspace (subspace $\tilde{P}$) on which $\chi$ acts nontrivially. Using conditions in Eqs.~(\ref{eq::S4}, \ref{eq::S5}) and ordering the basis such that the kernel vector is the last one, $\chi$, $\ket{v}$ and $\ket{q}$ have the following form in the new basis:
\begin{align}
    \chi_{\mathrm{b}} &= {\left(\begin{array}{l@{\quad}l}
    {\begin{array}{c}
      \scalemath{1.3}{\chi_{\tilde P}}_{\scalemath{0.6}{N_H-1 \times N_H -1}}
    \end{array}} & \scalemath{1.2}{0}_{\scalemath{0.6}{N_H-1 \times 1}} \\
    \\
    \ \ \scalemath{1.2}{0}_{\scalemath{0.6}{1 \times N_H-1}} & 0
  \end{array}\right)}_{\mathrm{b}}
  \notag\\[0.8ex]
  \ket{v}_{\mathrm{b}} &= {\left(\begin{array}{c}
    \scalemath{1.3}{0}_{\scalemath{0.6}{N_H-1 \times 1}}\\
    \\
    1
  \end{array}\right)}_{\mathrm{b}}, \quad
  \ket{q}_{\mathrm{b}} = {\left(\begin{array}{c}
    \scalemath{1.3}{q_{\tilde P}}_{\scalemath{0.6}{N_H-1\times 1}}\\
    \\
    0
  \end{array}\right)}_{\mathrm{b}},
\end{align}
where $\chi_{\tilde P}$ and $q_{\tilde P}$ are the matrix $\chi$ and vector $\ket{q}$ projected onto subspace $\tilde P$.
Subscript $b$ labels the new basis.
$\tilde{\chi}$ and its inverse are then
\begin{equation}
    \tilde\chi_{\mathrm{b}} = {\left(\begin{array}{l@{\quad}l}
    {\begin{array}{c}
      \scalemath{1.3}{\chi_{\tilde P}}
    \end{array}} & \scalemath{1.2}{0}\\
    \\
    \ \ \scalemath{1.2}{0} & 1
  \end{array}\right)}_{\mathrm{b}} \quad
  \tilde\chi^{-1}_{\mathrm{b}} = {\left(\begin{array}{l@{\quad}l}
    {\begin{array}{c}
      \scalemath{1.3}{\chi_{\tilde P}^{-1}}
    \end{array}} & \scalemath{1.2}{0}\\
    \\
    \ \ \scalemath{1.2}{0} & 1
  \end{array}\right)}_{\mathrm{b}}.
\end{equation}
Now one can easily see that the Ansatz in Eq.~\eqref{eq::regInv} is the solution to Eq.~\eqref{eq::S6}: $\chi_{\mathrm{b}} \tilde{\chi}_{\mathrm{b}}^{-1} \ket{q}_{\mathrm{b}} = \ket{q}_{\mathrm{b}}$. Moreover, it is also straightforward to show that the iterative conserved quantity, constructed from $\ket{w'}$, is traceless
\begin{align}
    \tr[\tilde{C}_k] &= \mathcal{N}_k^{-1} \sum_m w^{(k)}_m = \mathcal{N}_k^{-1} \braket{v|w'} =\notag\\
    &=\mathcal{N}_k^{-1} \bra{v}_{\mathrm{b}} \tilde{\chi}_{\mathrm{b}}^{-1} \ket{q}_{\mathrm{b}} = 0.
\end{align}

Let us note that the transformation to a new basis is useful for analytical arguments, however, to work with $\tilde\chi$ in practice, the original basis can be used with $\ket{v}\bra{v}=\frac{1}{N_H}\ket{1,...,1}\bra{1,...,1}$, as in Eq.~\eqref{eq::regInv}, expressed in that basis.

Finally we show that shifting the solution $\ket{w'}$ by $c\ket{v}$, Eq.~\eqref{eq::S6}, does not affect the steady state density matrix $\rho^{(k)}_{\boldsymbol{\tilde\lambda}}$ at the corresponding iterative step. 
Adding $c\ket{v}$ in the language of the iterative tGGE means adding a constant weight $c$ to each basis element $Q_m$, i.e.,
\begin{align}
\rho^{(k)}_{\boldsymbol{\tilde\lambda}}(c)
&\equiv \frac{e^{-\sum_{k'=0}^k \tilde\lambda_{k'} \mathcal{N}_{k'}^{-1}\sum_m (w_m^{(k')} + c) Q_m }}{\tr[{e^{-\sum_{k'=0}^k \tilde\lambda_{k'} \mathcal{N}_{k'}^{-1} \sum_m (w_m^{(k')} + c) Q_m }}]}\\
&= \frac{\sum_m e^{-\sum_{k'=0}^k \tilde\lambda_{k'}  \mathcal{N}_{k'}^{-1} (w_m^{(k')} + c) }Q_m }
{\sum_m[{e^{-\sum_{k'=0}^k \tilde\lambda_{k'} \mathcal{N}_{k'}^{-1} (w_m^{(k')} + c) }}]} \notag\\
&= \frac{e^{-\sum_{k'=0}^k \tilde\lambda_{k'} \mathcal{N}_{k'}^{-1} c} \sum_m e^{-\sum_{k'=0}^k \tilde\lambda_{k'}  \mathcal{N}_{k'}^{-1} w_m^{(k')} }Q_m }
{e^{-\sum_{k'=0}^k \tilde\lambda_{k'} \mathcal{N}_{k'}^{-1} c} \sum_m[e^{-\sum_{k'=0}^k \tilde\lambda_{k'} \mathcal{N}_{k'}^{-1} w_m^{(k')} }]}\notag \\
&=\rho^{(k)}_{\boldsymbol{\tilde\lambda}}. \notag
\end{align}
We see that the contribution from $c\ket{v}$ cancels out, so we can skip it in the first place. Similarly we could take a different shift $c \rightarrow c_{k'}$ for every iterative conserved quantity $\tilde C_{k'}$ and get the same result as above.

\subsection{Conservation laws of transverse Ising model} \label{app0}

For transverse field Ising model,
\begin{equation}
H_0 = \sum_i J\sigma^z_i \sigma^z_{i+1} + h_x \sigma^x_i,
\end{equation}
the conservation laws have the following form
\begin{align}\label{eq::isingC}
C_0&= H_0 \notag \\
C_2&=\sum_j 
J S^{zz}_{_{j,j+2}}
- h_x \sigma^y_{j}\sigma^y_{j+1} 
- h_x \hat\sigma^z_{j} \sigma^z_{j+1}
- J \sigma^x_{j} \notag \\
 C_{2\ell>2}&=  \sum_j 
J  S^{zz}_{_{j,j+\ell+1}} 
- h_x  S^{yy}_{_{j,j+\ell}}
- h_x  S^{zz}_{_{j,j+\ell}}
+J  S^{yy}_{_{j,j+\ell-1}} \notag \\
C_{2\ell-1}&= J \sum_j  S^{yz}_{_{j,j+\ell}} -  S^{zy}_{_{j,j+\ell}},
\end{align}
with $S^{\alpha\beta}_{i,j}=\sigma^\alpha_i \sigma^x_{i+1} \dots\sigma^x_{j-1} \sigma^\beta_{j}$. 
In our analysis we use $N_C$ most local ones from $H_0$ to $C_{N_C-1}$.

\subsection{Example II: Heisenberg model.}

\begin{figure}[t!]
\centerline{\includegraphics[width=\columnwidth]{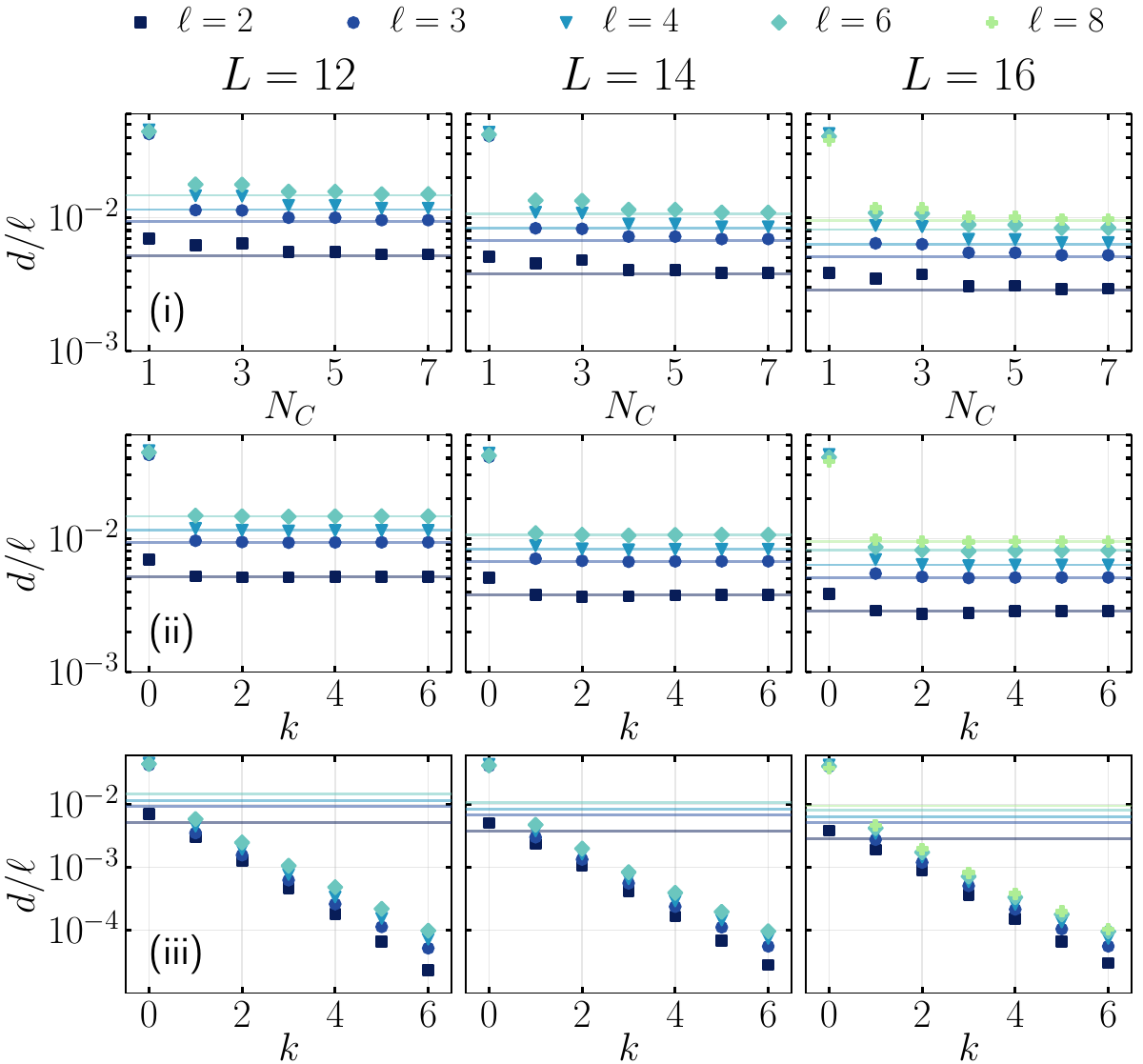}}
\caption{Scaled distances $d(\rho_{\boldsymbol{\lambda},\ell},\rho_{D,\ell})/\ell$, Eq.~\eqref{eq::dist}, between the reduced density matrices of support $\ell$ of the diagonal steady state ensemble, Eq.~\eqref{eq::diag}, and different truncated generalized Gibbs ensembles. Solid lines correspond to the tGGEs including $N_C^{\max}=8$ most local conserved quantities. From top to bottom we consider tGGEs with: (i)
$N_C$ local conserved quantities $C^{\text{loc}}_m, m=0,...,N_C-1$, 
(ii) iterative scheme after the $k$th step for basis $Q_m=C^{\text{loc}}_m, m=1,...,N_C^{\max}-1$, (iii) iterative scheme after the $k$th step for basis $Q_m=\ket{m}\bra{m}$. Parameters: Heisenberg model, system sizes $L=12,14,16$.}\label{supfig1}
\end{figure}

\new{Here, we perform a similar analysis to Fig.~1 in the main text for the paradigmatic interacting integrable model, the isotropic Heisenberg model,   
\begin{align}\label{eq::heis}
H_0 = \sum_i \sigma^x_i \sigma^x_{i+1} + \sigma^y_i \sigma^y_{i+1} + \sigma^z_i \sigma^z_{i+1},
\end{align}
and Lindblad operators 
\begin{equation} 
L_i = S_i^+S_{i+1}^- + i S_{i+1}^-S_{i+2}^{+}. \label{eq::heisL}
\end{equation}
They preserve magnetization; therefore, each magnetization sector has its unique steady state, and we consider below the largest zero magnetization sector.
Unlike the Ising model, the Heisenberg model has a few families of conserved operators: local extensive operators \cite{grabowski95} and quasi-local extensive operators \cite{ilievski15a,ilievski15,ilievski16}.}

Fig.~\ref{supfig1} shows the same analysis previously done for the Ising model. 
(i)~\nnew{Traditional tGGE: }The first row shows the distances for tGGE based on the first $N_C$ local conserved quantities, derived using the boost operator, $B = -i \sum_j j h_j$, $C^{\text{loc}}_{m+1} = [B,C^{\text{loc}}_m]$  and $C^{\text{loc}}_0 = H_0$, where $h_j$ are Hamiltonian densities $H_0 = \sum_j h_j$ \cite{grabowski95}. In this notation, $C^{\text{loc}}_1$ is the conserved energy current. 
For chosen Lindblads, approximate saturation in $d$ happens for a rather small $N_C$ number of local conserved quantities, therefore we perform our best calculation at $N_C^{\text{max}}=8$ local conserved quantities $C_m^{\text{loc}}$ to obtain $d_{s,\ell}$ denoted by solid lines.
(ii)~\nnew{Iterative tGGE with local basis: }The second row shows the distances for the iterative approach, using the basis of first $N_C^{\text{max}}=8$ local conserved operators $Q_m=C^{\text{loc}}_m, m=1,...,N_C^{\text{max}}-1$, excluding the Hamiltonian. The asymptotic distances $d_{s,\ell}$ (solid line) are approximately obtained after cca. $k=1$ iterative step.
(iii)~\nnew{Iterative tGGE with non-local basis: }The third row shows the distances for the iterative approach using the non-local basis $Q_m=\ket{m}\bra{m}$ of projectors onto the eigenstates of $H_0$. Already after the first iterative step, distances $d$ are smaller then for the tGGE including local conserved quantities because the non-local basis $Q_m$ captures also the impact of quasi-local conserved quantities of the Heisenberg model \cite{ilievski15,ilievski15a,ilievski16}. Partially, the reduction of $d$ in the non-local basis is again due to non-local contributions on small systems considered. \new{In the main text, we present a finite-size analysis of iterative conserved quantities created in this non-local basis, which removes the non-local contributions.}

\subsection{Iterative procedure for unitary perturbation} \label{appUni}

In case of (additional) unitary perturbation $H_1$, the dissipator $\cl{D}$ in Eq.~\eqref{eq::weight} should be replaced (or supplemented, if both types of perturbations occur simultaneously) by $\cl{L}_1 \hat{Q}\cl{L}_0^{-1} \hat{Q} \cl{L}_1$, i.e.,
\begin{align}
\tilde{C}_k &= 
\mathcal{N}_k^{-1}
\sum_m w^{(k)}_m Q_m, \label{eq::weightUni}\\
w^{(k)}_m&= - \sum_n \left(\chi_{(k-1)}^{-1}\right)_{mn} 
\tr[Q_n (\cl{D} + \cl{L}_1 \hat{Q}\cl{L}_0^{-1} \hat{Q} \cl{L}_1) \rho_{\boldsymbol{\tilde\lambda}}^{(k-1)}]  \notag
\end{align}
where 
$\cl{L}_j \rho= -i [H_j,\rho]$ for $j=0,1$, 
$\cl{L}_0^{-1} \rho= - \int_0^\infty dt e^{t \cl{L}_0} \rho$ and 
$\hat{Q} \rho =\mathbb{1} - \sum_{m} \ket{m}\bra{m} \bra{m} \rho \ket{m}$ is a super-projector orthogonal to the diagonal subspace, which ensures that the action of $\cl{L}_0^{-1}$ is not singular. Unlike for the dissipator, where the contribution to the flow in the GGE manifold occurs already in the first order, for the unitary perturbation the first nontrivial contribution happens only in the second order as $\tr[Q_n \cl{L}_1 \rho_{\boldsymbol{\tilde\lambda}}^{(k-1)}]=0$ due to cyclicity of the trace \cite{lange17,lenarcic18}.

\end{document}